# Noncentrosymmetric Triangular Magnet CaMnTeO$_6$: Strong Quantum Fluctuations and Role of s$^0$ vs. s$^2$ Electronic States in Competing Exchange Interactions

*Xudong Huai,[1] Emmanuel Acheampong,[1] Erich Delles,[1] Michał J. Winiarski,[2] Maurice Sorolla II,[3] Lila Nassar,[4] Mingli Liang,[5] Caleb Ramette,[6] Huiwen Ji,[6] Allen Scheie,[7,8] Stuart Calder,[8] Martin Mourigal,[4] Thao T. Tran[1]\**

Xudong Huai, Emmanuel Acheampong, Erich Delles, Thao T. Tran*
Department of Chemistry, Clemson University, Clemson, South Carolina 29634, United States
E-mail: thao@clemson.edu

Michał J. Winiarski
Applied Physics and Mathematics and Advanced Materials Center, Gdansk University of Technology, 80-233 Gdansk, Poland

Maurice Sorolla II
Institute of Chemistry, University of the Philippines Diliman, Quezon City, Philippines 1101

Lila Nassar, Martin Mourigal
School of Physics, Georgia Institute of Technology, Atlanta, Georgia 30332, United States

Mingli Liang
Department of Chemistry, University of Houston, Houston, Texas 77204, United States

Caleb Ramette, Huiwen Ji
Department of Materials Science and Engineering, University of Utah, Salt Lake City, Utah 84112, United States

Allen Scheie
Los Alamos National Laboratory, Los Alamos, New Mexico 87545, United States

Allen Scheie, Stuart Calder

Neutron Scattering Division, Oak Ridge National Laboratory, Oak Ridge, Tennessee 37830, United States

Abstract

Noncentrosymmetric triangular magnets offer a unique platform for realizing strong quantum fluctuations. However, designing these quantum materials remains an open challenge attributable to a knowledge gap in the tunability of competing exchange interactions at the atomic level. Here, we create a new noncentrosymmetric triangular $S = 3/2$ magnet $CaMnTeO_6$ based on careful chemical and physical considerations. The model material displays competing magnetic interactions and features nonlinear optical responses with the capability of generating coherent photons. The incommensurate magnetic ground state of $CaMnTeO_6$ with an unusually large spin rotation angle of 127°(1) indicates that the anisotropic interlayer exchange is strong and competing with the isotropic interlayer Heisenberg interaction. The moment of 1.39(1) $\mu B$, extracted from low-temperature heat capacity and neutron diffraction measurements, is only 46% of the expected value of the static moment 3 $\mu B$. This reduction indicates the presence of strong quantum fluctuations in the half-integer spin $S = 3/2$ $CaMnTeO_6$ magnet, which is rare. By comparing the spin-polarized band structure, chemical bonding, and physical properties of $AMnTeO_6$ (A = Ca, Sr, Pb), we demonstrate how quantum-chemical interpretation can illuminate insights into the fundamentals of magnetic exchange interactions, providing a powerful tool for modulating spin dynamics with atomically precise control.

## 1. Introduction

Control of competing magnetic states at the atomic level is a promising avenue to realize strong quantum fluctuations directly relevant to current challenges in developing novel paradigms for information technology.[1] Quantum fluctuations can enhance coherent quantum dynamics, a prerequisite for future solid-state quantum computing.[2] In frustrated magnets, competing magnetic states are degenerated or separated by small energy barriers.[3] This energy landscape gives rise to novel spin states and exotic dynamics, possibly with enhanced quantum fluctuations, but the manipulation of these competing magnetic states and their dynamics is difficult.[4] Magnetism, when combined with broken crystallographic inversion symmetry, gives rise to uniquely controllable micro- and macroscopic physical properties that are not possible for their centrosymmetric counterparts.[5] In addition, asymmetric Dzyaloshinskii-Moriya exchange can be stabilized and enhanced in noncentrosymmetric magnets in the presence of isotropic Heisenberg interactions, potentially leading to vortex-like spin states, associated nontrivial topology spin physics, and improved quantum fluctuations.[6] Recent efforts have focused on realizing noncentrosymmetric triangular-lattice magnets that simultaneously display nonlinear optical responses and appreciable quantum fluctuations.[7] However, a significant challenge with these systems has been poor control over chemical bonding and electronic modification under the strict constraints required for manipulating spin dynamics. Although antiferromagnetic (AFM) ordering typically removes the inversion symmetry of the electronic structure, it remains difficult to predict and synthetically target triangular-lattice spin systems that facilitate light-induced spin modulation and enhanced quantum effects. In essence, this stems from the significant conceptual barrier in predicting the scale of competing magnetic interactions from physical principles alone.

In this work, we take a step towards addressing this challenge by realizing a new noncentrosymmetric triangular-lattice magnet, $CaMnTeO_6$, that displays competing AFM-FM interactions and nonlinear optical response. The chemical bonding of this system, when placed in the context of related materials $AMnTeO_6$ (A = Sr, Pb) casts light on *how* and *why* the overlap of the interacting atomic wavefunctions determines their physical properties.[8] Three design parameters are important for $CaMnTeO_6$. The first is the careful choice of the Ca ($I = 0$), Mn ($I = 5/2$-, 100%), Te ($I = 0$), and O ($I = 0$) elements based on their nuclear spins and stable isotopes. The second is the integration of half-integer spin $S = 3/2$ of $Mn^{4+}$ into the triangular lattice formed by the noncentrosymmetric $TeO_6$ framework. The third is the placement of the $Ca^{2+}$ ion with the $s^0$ frontier orbital in between the triangular planes to study the influence of the A site on interlayer magnetic coupling. These design considerations are chosen to ultimately improve isotope purity and spin coherence time for noncentrosymmetric magnetic systems – a necessary step for integrating quantum materials into large-scale quantum device architectures.[9] We study the contributions of electron, spin, orbital, and phonon components of $CaMnTeO_6$ to its magnetic, optical, and thermomagnetic properties. We supplement these experiments with density functional theory (DFT) calculations on this model material and other relevant systems $AMTeO_6$ (A = Sr, Pb). $Sr^{2+}$ possesses a similar $s^0$ frontier electronic state to $Ca^{2+}$ but at higher energy ($5s^0$ vs. $4s^0$). $Pb^{2+}$ is very close to $Sr^{2+}$ in size while having lone-pair electrons $6s^2$. This systematic consideration allows us to determine how orbital overlap and electronic structure influence intralayer and interlayer exchange interactions (Table S1-4). [10]

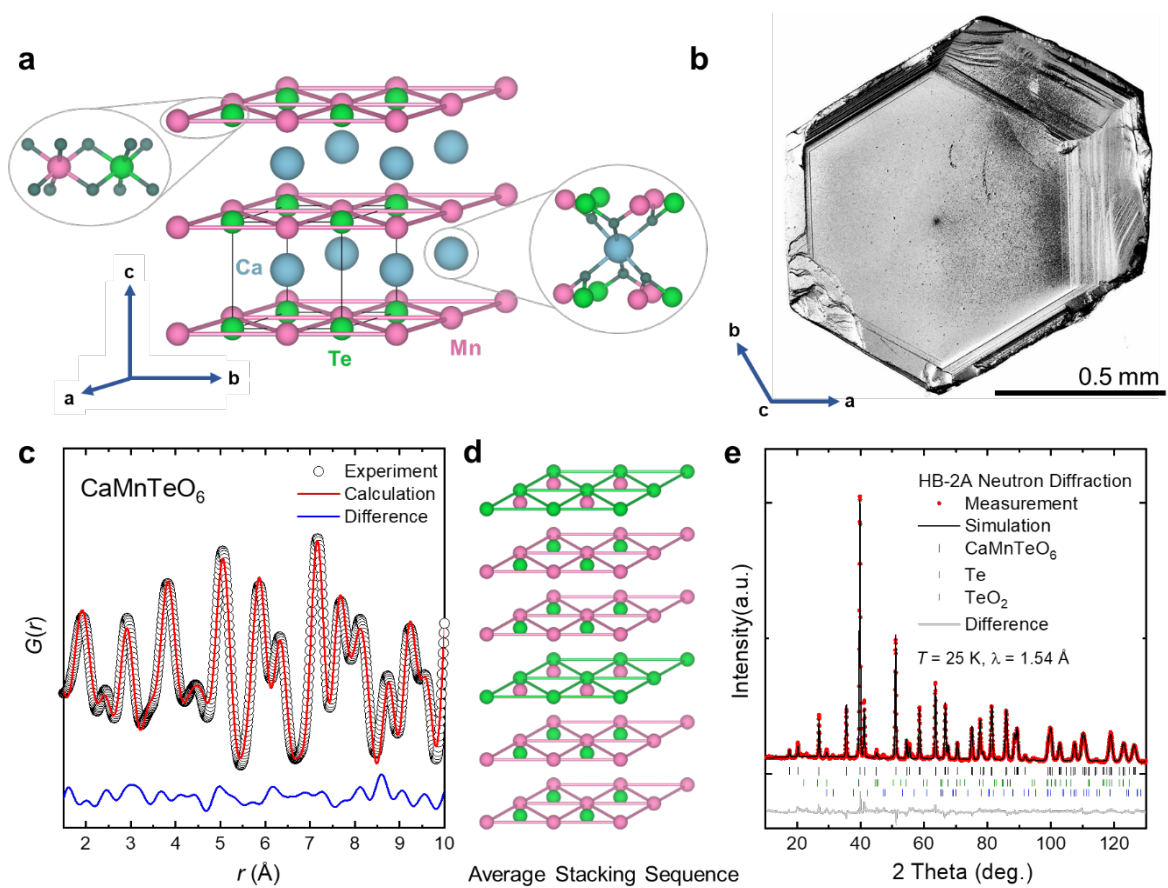

**Figure 1.** (a) Crystal structure of $CaMnTeO_6$ showing magnetic 2D triangular sublattice of $Mn^{4+}$ and layered crystal structure. (b) SEM image showing potential stacking fault along the *c*-axis. (c) Pair distribution function (PDF) data (black circle) and fitting (red). (d) Stacking fault model used for PDF analysis showing layered crystal structure shifting toward the [210] direction. (e) Refinement of HB-2A neutron diffraction data.

## 2. Results and Discussion

### 2.1. Crystal structure

The crystal structure of $CaMnTeO_6$ was determined by lab-based single crystal X-ray diffraction (SCXRD) and confirmed by reactor-based powder neutron diffraction (NPD, HB2A Oak Ridge National Laboratory), X-ray pair distribution function (PDF) analysis (Figure 1), and synchrotron powder XRD (11-BN Argonne National Laboratory (Figure S1a,b). The material crystallizes in the noncentrosymmetric chiral trigonal $P312$ space group and displays 2D triangular layers of $Mn^{4+}$ ($Te^{6+}$) separated by $Ca^{2+}$ ions through bridging oxygen atoms (Figure 1a). Each $Mn^{4+}$ ($Te^{6+}$) cation is coordinated to six O atoms in a distorted octahedral environment. The MnO$_6$ and TeO$_6$ octahedra are edge-sharing, forming the 2D triangular layer of Mn (Te) in the *ab*-plane. Along the *c*-axis direction, these layers are ionically bonded to the CaO$_6$ layer (Figure 1a).

The Mn and Te atoms switch their positions every 3 layers, yielding a stacking fault of the triangular layers of Mn (Te) along the *c*-axis direction. This imperfection can be attributed to similar radii of the Mn$^{4+}$ (0.53 Å) and Te$^{6+}$ (0.56 Å) cations.[11] The stacking fault features of CaMnTeO$_6$ were proved by the PDF analysis, a useful technique for characterizing local structures. It is worth noting that while this stacking fault may give an illusion of disordered Mn/Te, these atoms are, in fact, in ordered positions. The overall Mn/Te ratio and stacking fault were further confirmed by SEM-EDS and neutron diffraction experiment (Figure 1b-e, S2, S3). Such structural heterogeneities may play a nontrivial role in the spin environment of quantum magnets, and thus their magnetic excitations and ground states are similar to those in YbMgGaO$_4$ and KYbO$_2$ [12]

The intralayer Mn–Mn distance within the triangular lattice is 5.0607(4) Å, comparable to the interlayer Mn–Mn distance (5.0409(4) Å). In addition, the electronic structure of Mn$^{4+}$ is d$^3$ with three unpaired electrons populating the t$_{2g}$ state. These combined structural and electronic features may facilitate comparable exchange interactions in both intra- and interlayer in this material. Given this crystallographic structure, if the magnetic interactions of CaMnTeO$_6$ are solely captured by a 3D nearest-neighbor – Heisenberg model, a commensurate magnetic ground state is expected. However, the combination of broken inversion symmetry in the structure and Mn$^{4+}$ taking the $^4$A ground state in the $C_3$ crystal field with non-zero orbital angular momentum can facilitate off-diagonal anti-symmetric exchange interactions, potentially enabling an incommensurate magnetic state and enhancing spin fluctuation via competing interactions.[13]

## 2.2. Magnetic properties

The temperature-dependent magnetization of CaMnTeO$_6$ shows a subtle magnetic transition at $T_i$ = 9.7 K, determined by the minimum in d$M$/d$T$ vs $T$ (Figure 2a). The negative Curie-Weiss temperature $\Theta_{CW}$ of -25.5(1) K indicates dominant AFM interactions. The effective magnetic moment per formula unit extracted from the Curie-Weiss analysis is 3.9 (1) $\mu B$, which is very close to the expected value (3.88 $\mu B$) for the spin-only model of Mn$^{4+}$ ($S$ = 3/2). This excellent agreement confirms that there is one Mn cation per formula unit, consistent with the chemical compositions and structure discussed earlier. The $C_3$ local symmetry and the $^4$A ground state of the Mn magnetic cation are consistent with the electronic transitions observed in the UV-Vis-NIR spectrum analysis (Figure S4).

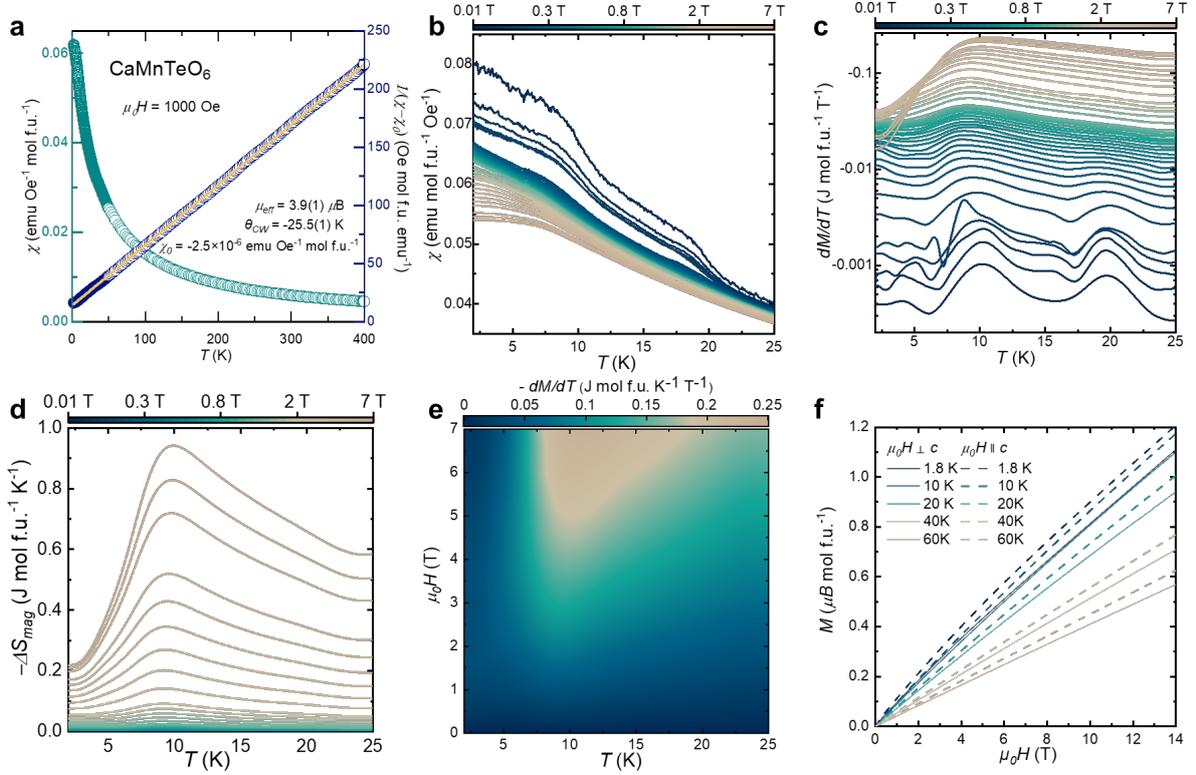

**Figure 2.** (a) Temperature dependent magnetic susceptibility under constant magnetic field (cyan) and Curie-Weiss analysis (blue). (b) Field-dependent magnetic susceptibility around transition temperature. (c) First derivative of magnetization with respect to temperature and magnetic field. (d) Isothermal magnetic entropy at each different magnetic field obtained by the integral of $dM/dT$ with respect to the magnetic field. (e) A map of $dM/dT = dS/dH$. (f) Orientation-dependent $M(H)$ curves at different temperatures.

In contrast to a conventional AFM or FM material, the magnetic ordering in the $\chi(T)=M(T)/H$ curve is very subtle and only confirmed by $dM/dT$ vs $T$ (Figure 2b-c). To assess whether a mixture of AFM and FM exchange interactions is present in the system, we study magnetoentropic signatures of CaMnTeO$_6$. Figure 2b shows how the magnetization of the material evolves as a function of temperature under different fields near the transition temperature. Two upturns at around $T$ = 20 K and 13 K are observed at low fields while the magnetic susceptibility appears to increase and nearly saturate at higher fields. The observed behavior in tandem with the negative sizable Curie-Weiss temperature $\Theta_{CW}$ imply appreciable competing AFM and FM interactions. The first derivative of magnetization with respect to temperature $dM/dT$ reveals that the transition temperature slightly goes up as the magnetic field increases (Figure 2c). This observation suggests that in the presence of applied magnetic fields, the spin entropy of FM coupling is decreased which is compensated by a rise in the lattice entropy of the material, resulting in an increase in the temperature. The isothermal magnetic entropy change is derived from the Maxwell relation (equation 1):

$$\left(\frac{dS}{dH}\right)_T = \left(\frac{dM}{dT}\right)_H \quad (1)$$

where $S$ is the total entropy, $H$ is the magnetic field, $M$ is the magnetization, and $T$ is the temperature. The $dM/dT$ map provides a complementary elucidation to the magnetic entropy (equation 2), $\Delta S_{mag}$ ($H$, $T$):

$$\Delta S_{mag}(H,T) = \int_0^H \left(\frac{dM}{dT}\right)_{H'} dH' \quad (2)$$

Figure 2d shows -$\Delta S_{mag}$ as a function of temperature under a series of applied magnetic fields 0.01 T ≤ $\mu_0 H$ ≤ 7 T. The sign of -$\Delta S_{mag}$ carries information about the nature of the phase transition, that is, a negative sign implies an AFM ordering while a positive sign represents an FM transition under applied fields. The value of the -$\Delta S_{mag}$ vs. $T$ curve is positive, suggesting the field-induced FM transition. The maximum of -$\Delta S_{mag}$ occurs around the magnetic phase transition temperature $T_i$ = 9.7 K and increases with an increase in an external magnetic field. The -$dM/dT$ = -$dS/dH$ map (Figure 2e) reveals diffuse ridges, implying field-driven phase transitions. From the -$dM/dT$ = -$dS/dH$ map, -$\Delta S_{mag}$ was calculated to be approximately 1.0 J mol$^{-1}$ K$^{-1}$. This value is in the same order of magnitude of that of Pr$_2$CuMnO$_6$.[10b, 14]

The results of the magnetoentropic mapping indicate competing AFM–FM interactions and field-induced FM transition. To answer whether CaMnTeO$_6$ manifests magnetic anisotropy, we turn to orientation-dependent $M(H)$ measurements on a single crystal at 0 T ≤ $\mu_0 H$ ≤ 14 T for both $\mu_0 H \perp c$ and $\mu_0 H \parallel c$ (Figure 2f) In both crystal directions, the magnetization does not saturate up to $\mu_0 H$ = 14 T. Nevertheless, there are noticeable differences in the $M(H)$ data in the $\mu_0 H \perp c$ and $\mu_0 H \parallel c$ directions. The magnetization in the $\mu_0 H \parallel c$ direction is greater than that in the $\mu_0 H \perp c$ at a given temperature, implying that there is a difference in the magnetic stiffness in the two orientations. This result also suggests that the Mn atoms within a triangular plane are antiferromagnetically coupled to one another (the intralayer interaction is AFM) whereas each Mn layer is asymmetrically correlated to adjacent layers (the interlayer interaction is anisotropic). The negative $\Theta_{CW}$ is indicative of the dominance of intralayer AFM interactions over the interlayer FM coupling. These anisotropic features are consistent with the magnetic susceptibility results discussed in Figure 2.

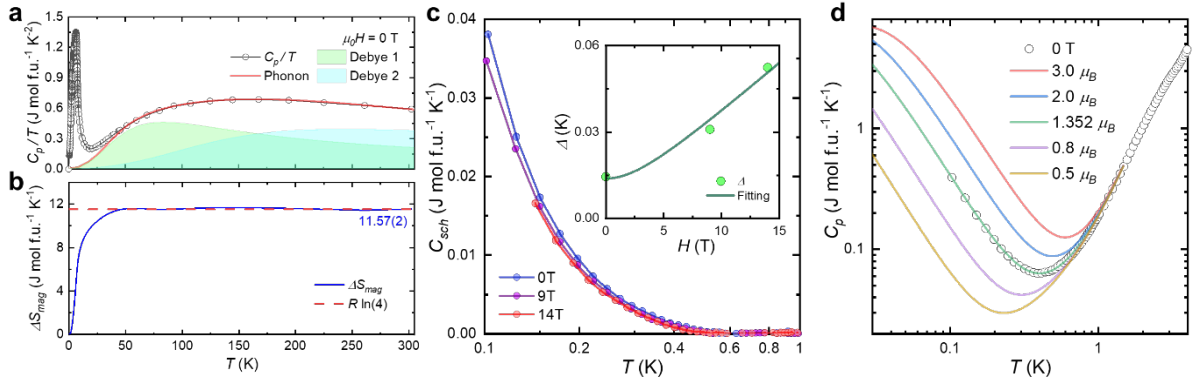

**Figure 3.** (a) Molar heat capacity over temperature ($C_p/T$) vs. temperature for CaMnTeO$_6$ at $\mu_0 H = 0$ T and calculated phonon. The anomaly is consistent with the magnetic phase transition of the material. (b) Magnetic entropy change $\Delta S_{mag} = 11.57(2)$ J mol$^{-1}$ K$^{-1}$ (blue line), consistent with the expected value of $S = 3/2$ spins (Rln 4) (red dash line). (c) Schottky heat capacity under different fields showing the Schottky effect is suppressed as increasing magnetic field. Inset: the Schottky gap extracted and the fitting shows that the gap is proportional to $\sqrt{(\langle \mu_s \rangle^2 + B^2)}$. (d) Heat capacity data and fitted models with different magnetic moments showing the best fit moment at $\mu_0 H = 0$ T is 1.352 $\mu_B$.

## 2.3. Thermomagnetic properties

To investigate the thermomagnetic properties of CaMnTeO$_6$, zero-field heat capacity measurements were conducted at $0.2$ K $\leq T \leq 300$ K (Figure 3a). An anomaly is observed in the specific heat $C_p/T$ vs. $T$ plot at $T = 8.5$ K, which is close to the transition temperature determined by magnetization measurements. This confirms the transition is magnetically driven. The magnetic entropy change (equation 3) $\Delta S$ from the transition can be calculated from:

$$\Delta S = \int_0^T \frac{C_v}{T} dT \tag{3}$$

where $C_v$ is the heat capacity at constant volume, which is approximated to be the same as $C_p$ (heat capacity at constant pressure) for solids at low temperatures, and $T$ is the temperature. Extracting the magnetic contribution to the specific heat is not trivial as the most direct nonmagnetic structural analog CaTiTeO$_6$ is unknown. Our attempts to create this new nonmagnetic phase were not successful. We thus chose to construct a phonon model to best describe the high-temperature specific heat data. The chosen model including two Debye modes (equation 4,5) is given as follows:

$$C_{Debye} = 9Nk \left(\frac{T}{T_D}\right)^3 \int_0^{T_D/T} \frac{x^3}{(e^x - 1)} dx \tag{4}$$

$$\frac{C_p}{T} = \frac{C_{\text{Debye}(1)}}{T} + \frac{C_{\text{Debye}(2)}}{T} \tag{5}$$

where $N$ is number of atoms, k is Boltzmann's constant, $x$ is the phase parameter ($\hbar\omega/k_B$), and $T_D$ is the Debye temperature. When fitting phonons in heat capacity, we commonly use Debye and Einstein models which describe acoustic and optic phonons. No Einstein mode was included since there was no characteristic $T$ max in the $C_p/T^3$ vs. $T$ plot (Figure S7). A combination of two Debye models fitted the experimental data well and yielded physical oscillator terms that added up to the total number of atoms in the formula unit. The criteria for our choice of the appropriate two-Debye model are based on the resulting good fit and physical oscillator terms. It is logically rationalized by the two subunits: the phonon modes of the triangular framework containing the magnetic Mn cation layer (i) are expected to be energetically distinct from those associated with the nonmagnetic sublattices (ii). The model parameters from the least-squares refinement to the specific heat data are summarized in Table S5. The total oscillator strength is 9.3(2), consistent with the expected value of 9 which is the total number of atoms per formula unit in $CaMnTeO_6$. After subtracting the phonon contribution, the change in entropy corresponding to the magnetic order was estimated to be 11.57(2) J mol f.u.$^{-1}$ K$^{-1}$, comparable to the expected recovery of $\Delta S_{mag} = R\ln(2S+1) = R\ln(4) = 11.5$ J mol f.u.$^{-1}$ K$^{-1}$ (Figure 3b). This matched entropy change suggests an absence of classical disorder.

To know more about potential quantum spin fluctuations, temperature-dependent heat capacity is measured in the dilution refrigerator region (0.1 K < $T$ < 1 K). The observed Schottky anomaly can only be attributed to the nuclear spin because of the energy scale under which it was observed.[15] To obtain a pure nuclear Schottky contribution, the electronic contribution ($C_e$) was subtracted from the measured heat capacity data. We estimated the electronic contribution under this temperature using the model $C_e = AT + BT^C$; where $A$, $B$, and $C$ are constants, and $C$ provides insight into the dimension of the spin wave. Additionally, we did not consider the phonon contribution, as it should be negligible at such low temperatures.

The resulting nuclear heat capacity data (Figure 3c) clearly showed the presence of the Schottky effect under different magnetic fields. The Schottky effect at 0 T arises from the magnetic field generated by the magnetic moment of the electronic spins ($\langle\mu_s\rangle$) on the nuclei. When an external field ($\boldsymbol{B}$) is applied, the onset of the Schottky peak appears to move to slightly lower temperature. To quantify the Schottky gap, a two-level model is used. Ideally, a complete model would include the 18-state hyperfine coupled state manifold from $S = 3/2$, $I =$ 5/2 (Figure S8). As the data only show a small upturn (a tail of a characteristic Schottky anomaly), all states have been populated at the temperature at which the data were analyzed. In other words, we can only extract the highest energy gap. Moreover, the shape of a two-level Schottky model is not very different from that of the complete hyperfine coupling model. Thus, the Schottky gap is fitted from $C_{sch}$ vs. $T$ curve. The result (Figure 3c) showed that the Schottky gap is proportional to $\sqrt{(\langle \mu_s \rangle^2 + \boldsymbol{B}^2)}$.

In addition, to estimate the local static moment in CaMnTeO$_6$, we construct a nuclear hyperfine model using the Mn$^{4+}$ hyperfine coupling constants[16] and the $I = 5/2$ nuclear state of naturally occurring Mn and the Hamiltonian (equation 6)

$$\mathcal{H} = A\langle m \rangle I_Z \tag{6}$$

where $\langle m \rangle$ is the static electronic magnetic moment, $A$ is the nuclear hyperfine coupling constant, and $I_Z$ is the nuclear angular momentum. We then calculated heat capacity via the derivative of free energy[17]. Modeling heat capacity below 1 K as a nuclear Schottky anomaly plus a fitted power law for the magnetic heat capacity, we find the local static moment value is very well constrained by the fit, as shown in Figure 3d. The best-fitted moment 1.352 $\mu B$ is approximately 46% compared with the static magnetic moment calculated from the free-ion $g$-factor (equation 7):

$$m = g_0 \cdot S \approx 2 * \frac{3}{2} = 3.00 \ \mu_B \tag{7}$$

To dig deeper into whether a possible reduction of the effective $g$-factor at low temperatures influences the analysis, the $g$-factor of CaMnTeO$_6$ was estimated from a point-charge model as $g_{cal} = 1.379$ using PyCrystalField software[18]. The deviation of the $g$-factor from 2.002 could be attributed to the covalency of chemical boning and the anisotropy in this system. Using this $g_{cal}$ in the equation to estimate the static magnetic moment, one would expect 1.261 * 3/2 = 1.892 $\mu B$. In this case, the fitted moment of 1.352 $\mu B$ still falls short (71%) of the expected static magnetic moment, signaling quantum fluctuations in the ground state.

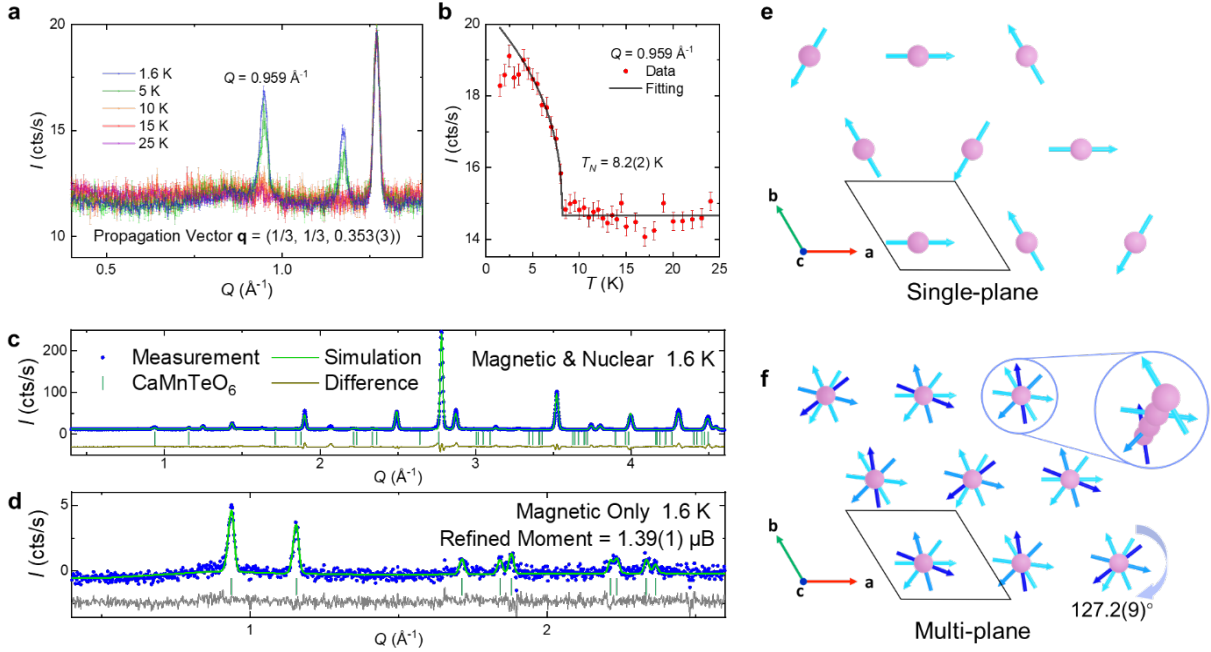

**Figure 4.** (a) HB-2A powder neutron diffraction with a wavelength of 2.41 Å of $CaMnTeO_6$ under different temperatures showing new Bragg peaks with the propagation vector **q** = (1/3, 1/3, 0.353(3)). (b) The intensity of the most intense magnetic Bragg peak at different temperatures showing the Neel temperature of 8.2(2) K. (c) Nuclear and magnetic structure refinement. (d) Magnetic diffraction refinement showing the proposed magnetic structure fits the data well. (e,f) Refined magnetic structure showing 120° classical Neel ground state on each *ab*-plane and incommensuration along the *c*-axis.

## 2.4. Magnetic structure

To understand the magnetic structure and confirm the relatively small magnetic moment observed in low-temperature heat capacity measurement in $CaMnTeO_6$, we turn to neutron powder diffraction at low temperatures. The neutron diffraction confirmed that the ordering temperature is at 8.2(2) K (Figure 4a-b). Refinement of the nuclear and magnetic structure (Figure 4c-d) was performed using the Fullprof software[19] based on the average structure. We isolated the magnetic Bragg peaks by subtracting 20 K data (above $T_i$) from the 1.6 K data (below $T_i$). The magnetic Bragg peaks can be indexed by a single propagation vector **q** = (1/3, 1/3, 0.353(3)), indicating commensurate magnetism in the *ab*-plane and incommensurate magnetism along the *c*-axis. Using irreducible representation analysis (see methods section) we find the diffraction pattern matches that of a coplanar spiral structure (Figure 4e). The coplanar 120° ground state magnetic structure is the ground state magnetic order of the Heisenberg antiferromagnet[20]. The classical triangular lattice phase diagram has been thoroughly studied theoretically[21], and a 120° ordered structure indicates that the dominant in-plane exchange in $CaMnTeO_6$ is the nearest neighbor Heisenberg. One peculiar note about the structure is that each triangular lattice plane is within the classical 120° Neel manifold, but each spin is rotated 127(1)° from the spin beneath it (Figure 4f), leading to an incommensurate spiral along the *c*-axis. This means the rotation angle is 53(1)°, less than an ideal AFM-stacking along the *c*-axis, which is unusual for a 3d ion. Two proposed contributions are Dzyaloshinskii–Moriya (DM) interaction and stacking faults. The anisotropic DM interaction[22] can be expressed by the equation 8:

$$H_{i,j}^{(DM)} = \boldsymbol{D}_{ij} \cdot (\boldsymbol{S}_i \times \boldsymbol{S}_j) \tag{8}$$

where $\boldsymbol{D}_{ij}$ is a vector determined by the symmetry of the lattice between spins *i* and *j*. For the nearest neighbors out of the plane, $\boldsymbol{D}_{ij}$ is constrained to be along the *c*-axis due to a lack of inversion symmetry and a three-fold rotation axis, which in competition with a Heisenberg exchange *J* would tend to produce an incommensurate spiral magnetic ground state[6e]. Under this hypothesis, the observed rotation angle indicates a ratio $D/J = \tan(2\pi Qc) = 1.32(5)$, indicating an antiferromagnetic interplane Heisenberg interaction *J*, and the anisotropic inter-plane exchange is strongly competing with the isotropic inter-plane Heisenberg exchange in $CaMnTeO_6$. Another explanation for this incommensuration is stacking faults because the incommensuration vector, 0.353 (3), is close to 1/3 (given by pure stacking fault). This implies that the stacking fault structure where the Mn and Te switch their position every three layers participates in stabilizing the incommensurate structure. It has also been proved by previous research that stacking fault/disorder/doping can result in incommensurate magnetic ordering[23]. Still, the propagation vector along *c*-axis differing from 1/3 suggests that stacking faults are not the only factor causing incommensurate magnetism. Besides the DM interaction, there can be other possible explanations such as anisotropic inter-plane interactions and biquadratic exchange interactions, but their contributions are expected to be small in 3d ions. Nevertheless, the magnetic structure tells us that $CaMnTeO_6$ is dominated by in-plane Heisenberg antiferromagnetism.

The refined magnetic moment determined by the intensity of magnetic diffraction under base temperature is 1.39(1) $\mu B$, ~ 46% of the expected value of the static moment 3 $\mu B$, which agrees very well with the fitted moment from the nuclear hyperfine heat capacity 1.352 $\mu B$. This indicates that the $Mn^{4+}$ local static moments all participate in the global 120° magnetic ordered ground state, and there is no static spin disorder. The agreement of the reduced moment proves the presence of strong quantum fluctuations in $CaMnTeO_6$. While similar quantum phenomena have been observed in other triangular-lattice magnets, this realization in the relatively large spin *S* = 3/2 system is rare.[24]

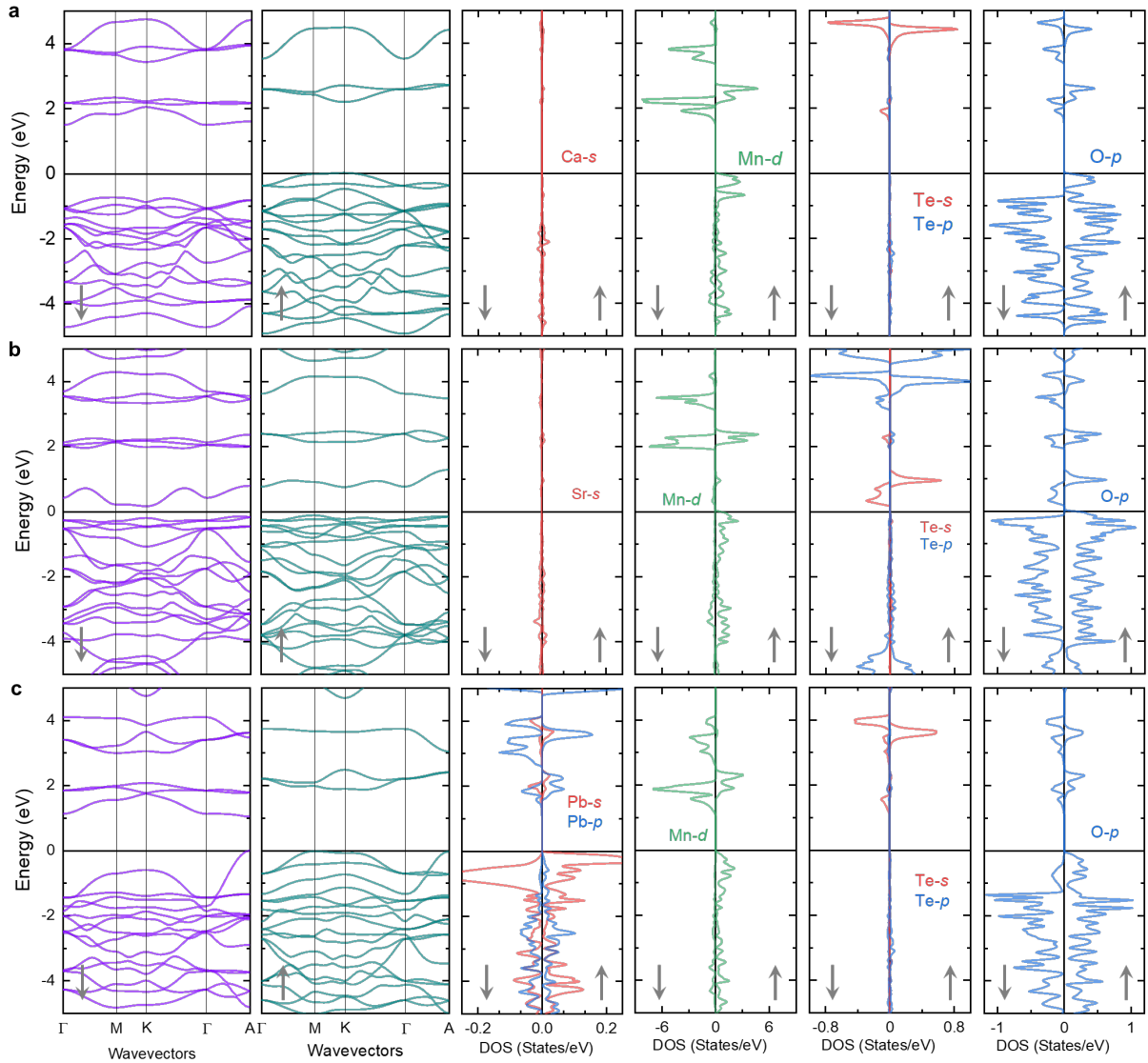

**Figure 5.** Spin-polarized band structure showing diffused bands around the Fermi level and spin polarized density of states (DOS) of (a) CaMnTeO$_6$, (b) SrMnTeO$_6$, PbMnTeO$_6$ showing polarized Mn, Te, A, and O.

## 2.5. Density functional theory calculations

*2.5.1 Spin-polarized band structure and density of states*

To gain more insight into how the electronic structure of AMnTeO$_6$ (A = Ca, Sr, Pb) noncentrosymmetric magnets manifest in triangular intralayer and interlayer coupling, full-potential spin-polarized DFT calculations were performed using WIEN2k (Figure 5). The results clearly demonstrate some common features of AMnTeO$_6$. The bands around the Fermi level are diffuse, suggesting good overlapping between the Mn-*d*, O-*p* states, and directional bonding features (Figure 5). The spins of the Mn-*d* states are polarized, and further polarize the O-*p*, Te-*s*/*p*, and A-*s* (A = Ca, Sr, Pb) states. Taken together, the directional bonding characters and spin polarization support the magnetic properties of AMnTeO$_6$.[10a, 10h]. The electronic band structure also helps explain the magnetic anisotropy of CaMnTeO$_6$ observed in the aforementioned physical properties and neutron experiment. However, the level of band diffusion and the contribution of the s-states of the A site around the Fermi level ($E_F$) are different. The s-states of Ca and Sr are fully oxidized, and their density of states (DOS) contribute mostly at low energy well below $E_F$. On the contrary, the lone-pair electrons (s$^2$) of Pb contribute significantly to the DOS around $E_F$. This departure in the DOS and band structure of AMnTeO$_6$ is expected to show up in magnetic properties of the materials, especially intralayer vs. interlayer coupling.

*2.5.2 Spin density map*

The intralayer and interlayer exchange pathways were mapped out by using the spin density map ($\rho_{up} - \rho_{down}$) and projected on selected lattice planes (Figure 6). Figure 6 a-b highlights the spin polarization on the [001] plane. The Mn-*d* magnetic spins polarize the spin density of the O-*p* states, which then polarize the Te site, forming AFM intralayer exchange interactions within the *ab*-plane through Mn-O-Te-O-Mn. The O-*p* spins in PbMnTeO$_6$ are more polarized by the Mn-*d* states from the e$_g$ orbitals than those in the Ca and Sr materials. Figure 6 c-d depicts the spin polarization on the [100] and [110] plane, respectively. It is apparent that the polarized O-*p* spins, generated by the magnetic density on Mn sites, induce appreciable spin polarization on the A site, forming interlayer exchange pathways along the *c*-axis direction through Mn-O-A-O-Mn. The spin polarization on the A site increases from Ca (4$s^0$) to Sr (5$s^0$) and Pb (6$s^2$).

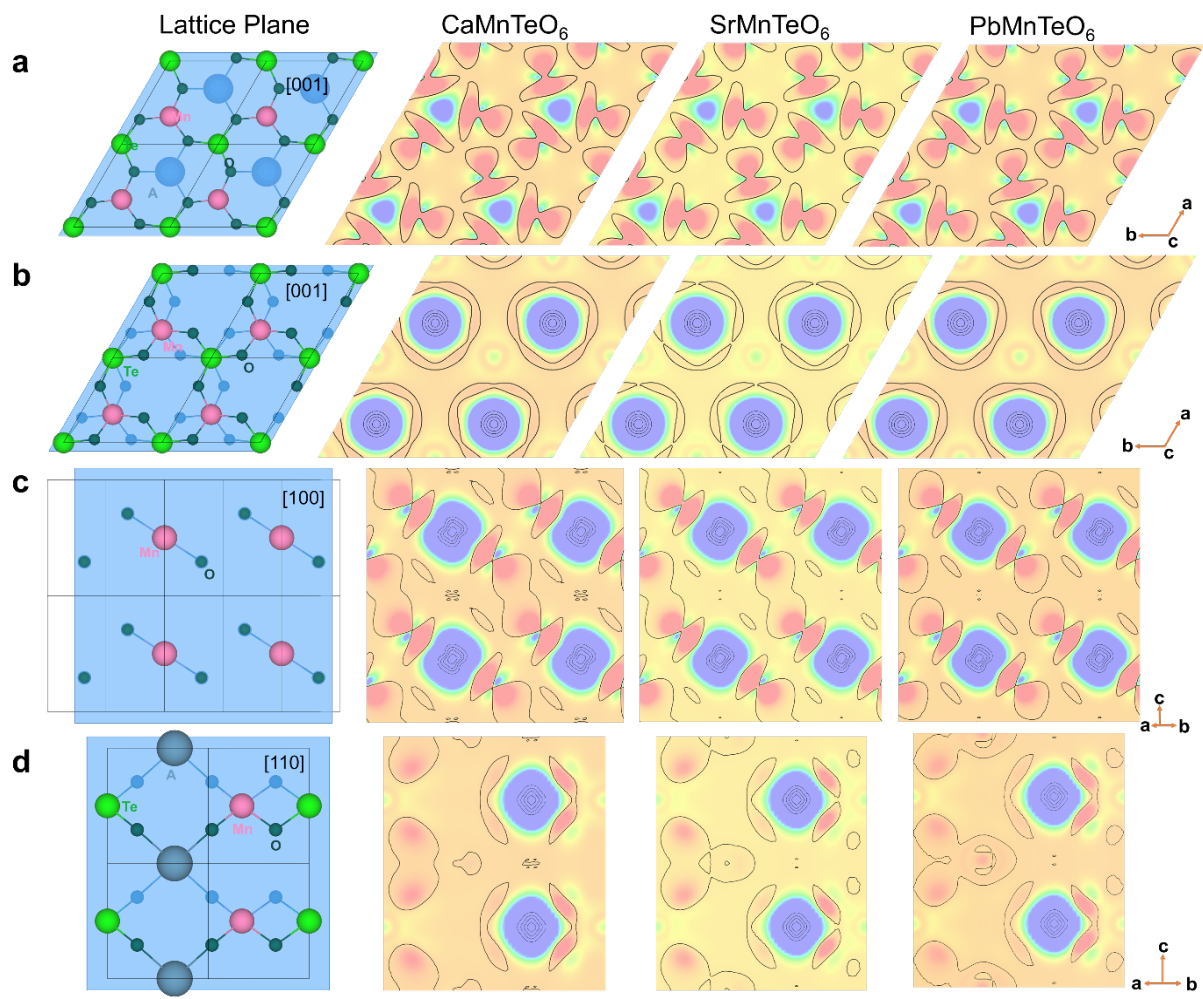

**Figure 6**: Spin density map of $AMnTeO_6$ (A = Ca, Sr, Pb) (a) [001] plane cutting through the O layer between Mn and Te sites. (b) [001] plane on the Mn/Te layer. (c) [100] plane cutting through Mn-O bonds. (d) [110] plane cutting through Mn and A sites.

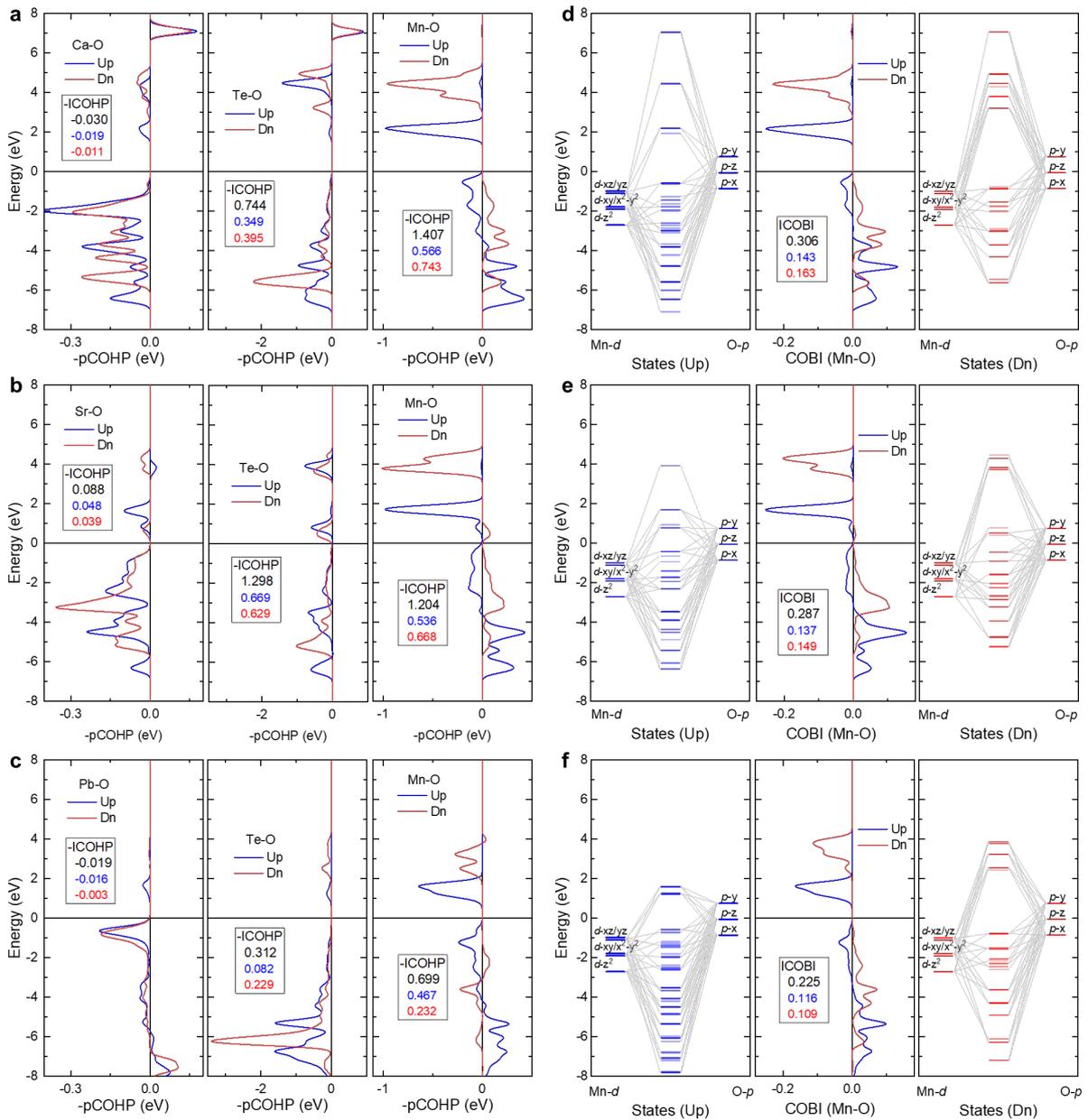

**Figure 7.** (a-c) Projected crystal orbital Hamilton populations (-pCOHP) with their integrated value (-ICOHP: Total (black), Spin-Up (blue), Spin-Dn (red)) for (a) CaMnTeO$_6$, (b) SrMnTeO$_6$, and (c) PbMnTeO$_6$. (d-f) Crystal orbital bond index (COBI) with their integrated value (ICOBI) up to $E_F$ and fragment crystal orbital diagrams for (d) CaMnTeO$_6$, (e) SrMnTeO$_6$, and (f) PbMnTeO$_6$.

### 2.5.3. Chemical Bonding

To understand how the bonding conditions of AMnTeO$_6$ (A = Ca, Sr, Pb) influence their competing magnetic interactions, additional pseudopotential DFT calculations were performed using the projector-augmented wave (PAW) method as implemented in the Quantum Espresso software,[25] and then projected into a linear combination of atomic orbitals (LCAO) based representation using Local Orbital Basis Suite Towards Electronic-Structure Reconstruction (LOBSTER)[26] program.[27] The DFT results without spin-polarization show metallic character in the DOS of AMnTeO$_6$, which is not true, and sizeable antibonding at around $E_F$ (Figure S10). This proves that a phase transition ought to occur, either structurally or magnetically, to lower the symmetry, thus stabilizing the system. Since there is no structural phase transition observed, undergoing a magnetic phase transition is a means through which the electronic instability of AMnTeO$_6$ is alleviated. The spin-polarized DOS curves from pseudopotential (Figure S11) are parallel to those from full-potential (Figure 5), showing the comparable contribution of the Mn-$d$ and Te-$s$/$p$ states in AMnTeO$_6$ and the contrast in the participation of the s$^0$ (Ca$^{2+}$ and Sr$^{2+}$) and s$^2$ (Pb$^{2+}$) frontier orbitals of the A-site. In addition, zero DOS and crystal orbital Hamilton population (COHP) at $E_F$ suggest that the materials stabilize themselves by displaying their magnetic properties. Although the DOS is helpful in describing the state contribution, it does not contain the phase information of the orbitals involved in the overlap of the wavefunctions (constructive vs. destructive interference). LOBSTER program, developed by Dronskowski et al., enabled us to reconstruct the PAW wavefunctions to extract the vital phase information. The projected crystal orbital Hamilton population (-pCOHP)[28] curves indicate nonbonding for the Ca-O and Sr-O bonds, antibonding (-pCOHP < 0) for the Pb-O bonds, and nonbonding-weak antibonding for the Te-O bonds, and antibonding for the Mn-O bonds in the vicinity of $E_F$ (Figure 7a-c). The spin-up (majority spin) and spin-down (minority spin) COHP curves of A-O and Te-O are similar in shape but differ in size. On the other hand, the Mn-O spin-up and spin-down COHP curves in AMnTeO$_6$ have distinct shapes and significant shifts in energy. The spin-up states see a larger nuclear charge, thereby decreasing in energy. The spin-down states, on a contrary, experience more effective shielding from the nucleus and raise in energy. These changes lead to divergence in the spatial extents of the two sets of spin-up and spin-down sublattices. These inequivalent spin sublattices reduces the electronic symmetry of AMnTeO$_6$, thus stabilizing the system and giving rise to magnetism. The total integrated Mn-O COHPs (ICOHPs) are 1.309, 1.207, and 0.698 eV per bond for the Ca, Sr, and Pb material, respectively, implying that the strongest Mn-O bonding in CaMnTeO$_6$. Similar features are also observed in crystal orbital bond index (COBI) analysis that describes pairwise Mn-O interactions (Figure 7d-e)[29]. The integrated Mn-O COBI (ICOBI) values are 0.305, 0.287, and 0.225 for the Ca, Sr, and Pb material, respectively, giving a hint that the Mn-O bonding character in CaMnTeO$_6$ is the most covalent among those in the series. To further compare the bonding situation of the Mn-O bonds in AMnTeO$_6$, the fragment crystal orbital (FCO) diagrams were constructed from the combination of the Mn-d and O-p DOS curves and the two-center Mn-O COBI. The FCO analysis describes the pairwise Mn-O interactions, resembling a classic molecular orbital approach. Overall, the spin-up states in the FCO diagrams are lower in energy than the spin-down states, consistent with the COHP and COBI results. Nevertheless, a closer look reveals that both the spin-up and spin-down states of Mn-O in $CaMnTeO_6$ are most diffuse in the series while those in the Pb material are most contracted. This proves how quantum-chemical interference phenomena wherein the atomic wavefunctions in $AMnTeO_6$ interact constructively (bonding) or destructively (antibonding) manifest in their physical properties. While the FCO and two-center COBI analysis allowed us to understand the pairwise interactions, multicenter COBI consideration is essential to dive deeper into the intralayer ($J_1$) and interlayer ($J_2$) interactions in the triangular magnets. Three-center COBI was calculated for Mn-O-Te and Mn-O-A as a representation for the Mn-O-Te-O-Mn intralayer and Mn-O-A-O-Mn interlayer interaction, respectively (Figure 8). The negative ICOBI suggested a multicenter interaction, thus implying super-super exchange coupling occurring both within the triangular layers and between the layers.

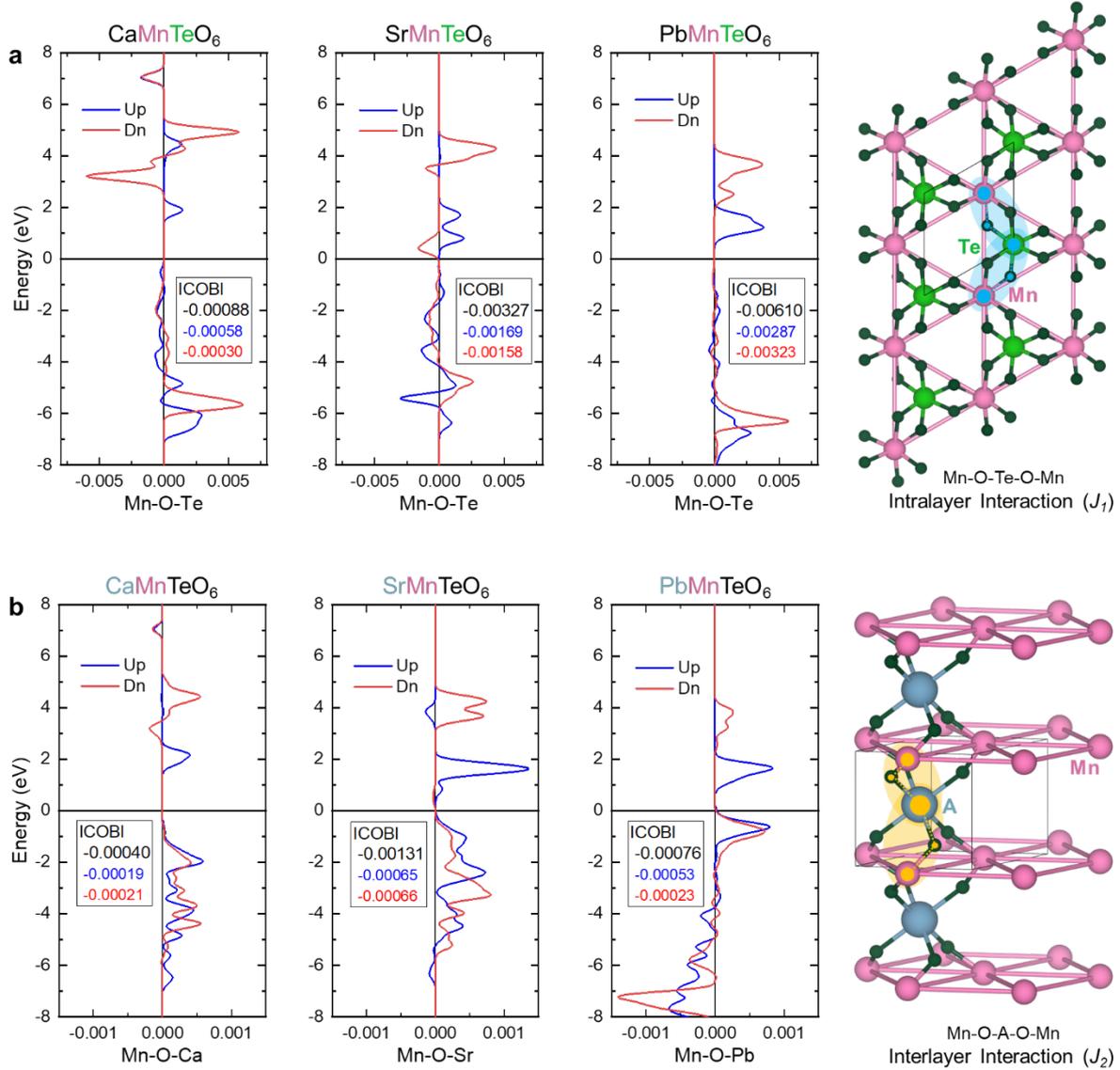

**Figure 8.** Three-center crystal orbital bond index (COBI) plots for CaMnTeO$_6$, SrMnTeO$_6$, and PbMnTeO$_6$, showing the super-super exchange interaction pathway (a) intralayer ($J_1$) and (b) interlayer ($J_2$).

*2.5.4. Magnetic exchange interactions*

To estimate $J_1$ and $J_2$ exchange interactions in AMnTeO$_6$, we applied the Green's function method by using the Wannier functions formalism through DFT and Heisenberg model.[30] This approach results in all the exchange interactions from the calculation of a magnetic configuration while proving insights into orbital contributions to the total exchange coupling. The most competing intralayer and interlayer exchange interactions were identified for CaMnTeO$_6$, $J_1$ = -9 K (AFM) and $J_2$ = 9 K (FM). This result is in harmony with the magnetic ground state deduced from neutron experiments. SrMnTeO$_6$ displays AFM interactions both within and between layers ($J_1$ = -16 K and $J_2$ = -29 K) while PbMnTeO$_6$ features FM exchange constants ($J_1$ = 133 K and $J_2$ = 31 K), consistent with their reported magnetic properties.[10a],[10f] It is worth noting that the $J_1$ and $J_2$ values of the Sr and Pb systems are quite different in size and not nearly as competitive as those of the Ca material.

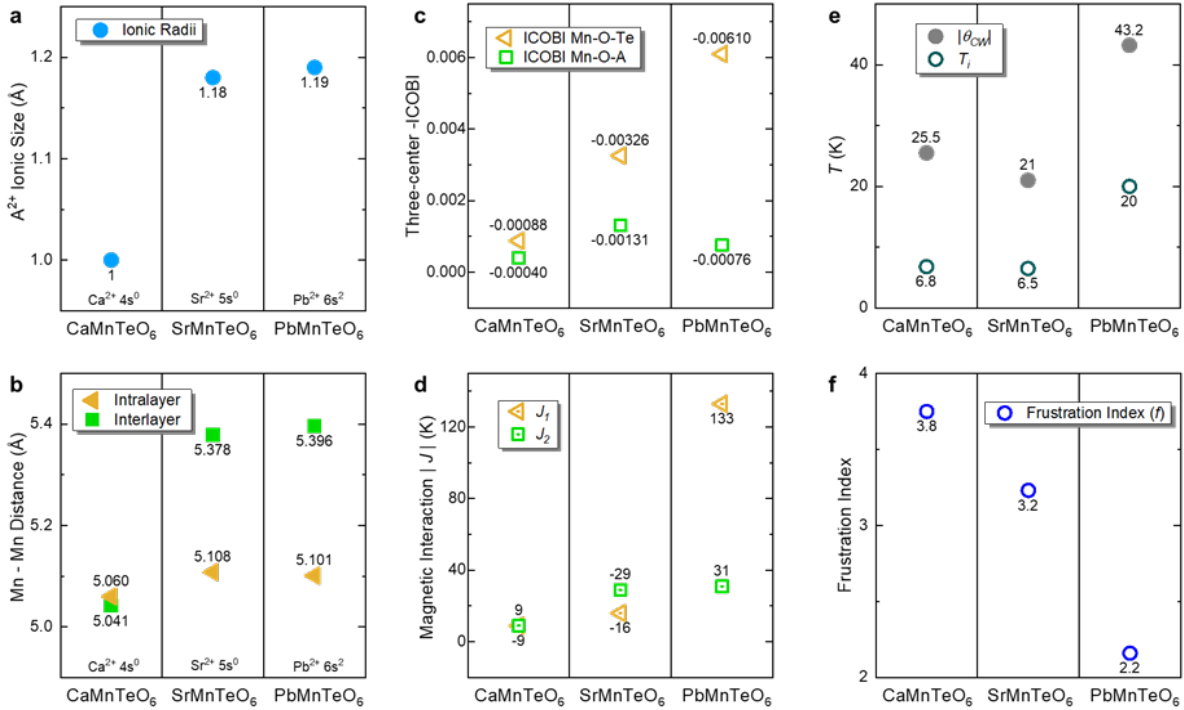

**Figure 9.** (a) Ionic radii of the A sites (A = Ca, Sr, and Pb). (b) Mn-Mn distances in AMnTeO$_6$. (c) Integrated three-center crystal orbital bond index (ICOBI) up to $E_F$ for AMnTeO$_6$ showing the multicenter intralayer and Mn-O-A interlayer interactions. (d) Calculated intralayer $J_1$ and interlayer $J_2$ magnetic interactions for AMnTeO$_6$ showing the comparison of the Mn-Mn Heisenberg exchange constants. (e) Curie-Weiss temperature ($\theta_{CW}$) and ordering temperature ($T_i$). (f) Frustration index ($f$) of AMnTeO$_6$.

Figure 9 depicts how the electronic structures and bonding analysis of AMnTeO$_6$ can be tied to their crystal lattices and physical properties. By substituting the A site with $Ca^{2+}$ – a smaller cation having the $4s^0$ frontier electronic state, the title material CaMnTeO$_6$ features shortest intralayer and interlayer distances and most comparable multicenter ICOBI for intralayer and interlayer interactions, and thus the most competing Heisenberg $J_1$ and $J_2$ exchange constants. This enhances competing AFM exchange coupling within the triangular layer and FM between the layers, lowering the ordering temperature (Figure S12) and facilitating the magnetic frustration ($f = |\theta_{cw}|/T_i = 25.5/6.8 \approx 3.8$ – highest frustration index in the series). SrMnTeO$_6$ exhibits similar $T_i$ but lower frustration index ($f = |\theta_{cw}|/T_i = 21/6.5 \approx$ 3.2), attributable to the $5s^0$ frontier electronic state. When Sr is replaced by Pb, which is comparable in size but different in valence electrons ($s^0$ vs. $s^2$), the Mn–Mn intralayer and interlayer distances in AMnTeO$_6$ (A = Sr, Pb) are very close. Nevertheless, the multicenter ICOBI interactions and exchange constants within and between layers in the Pb material are significantly different. This observation suggests that less competing $J_1$ and $J_2$, giving rise to highest ordering temperature and smallest frustration index ($f = |\Theta_{cw}|/T_i = 43.2/20 \approx 2.2$). In systems where competing exchange interactions have similar energy scales, care should be taken when using this qualitative measure of magnetic frustration. The systematic consideration of AMnTeO$_6$ (A = Ca, Sr, Pb), which exhibits some common features in exchange pathways, can justify the aforementioned interpretation.

## 3. Conclusions

These results demonstrate a framework to design, modify, and ponder noncentrosymmetric triangular magnets for atomically controlling competing magnetic states. Judicious considerations of isotope purity, nuclear and electronic spins, lattice symmetry, frontier orbitals and electronic states for these systems can result in realization of new physical phenomena. With this design principle in mind, we create a previously untapped noncentrosymmetric triangular $S = 3/2$ magnet, CaMnTeO$_6$, that features competing intralayer and interlayer magnetic interactions while displaying the capability of generating coherent photons. We find that the model triangular magnet possesses an incommensurate spiral magnetic ground state with 120° Neel manifold in the *ab*-plane and a spin rotation angle of 127°(1) along the *c*-axis. This spin rotation angle is significantly larger compared to other incommensurate systems, indicating that an anisotropic interlayer exchange is strongly competing with the isotropic interlayer Heisenberg interaction. The observation suggests that anisotropic interactions, stacking fault, and orbital overlapping between the layers factor in the central Hamiltonian. The consistent, reduced moment, extracted from both the low-temperature electro-nuclear heat capacity and the neutron diffraction data, reveals strong quantum fluctuations in CaMnTeO$_6$, which is rare for $S = 3/2$ systems. This can be attributed to the broken spatial symmetry, the Mn-O covalency, and the effectiveness of the overlap of the interacting atomic orbitals within and between the triangular layers in this material, resulting in comparable intralayer and interlayer exchange interactions.

By contrasting the chemical bonding and magnetic properties of AMnTeO$_6$ (A = Ca ($4s^0$), Sr ($5s^0$), Pb($6s^2$)), we connect quantum-mechanical interference phenomenon to the underlying physics of competing exchange interactions. CaMnTeO$_6$ displays the most Mn-O covalent bonding character, the most dispersed spin-up and spin-down Mn-O states, and the most comparable multi-center ICOBI and competing *J* coupling constants for intralayer and interlayer exchange pathways in the series. This is in harmony with enhanced magnetic frustration and increased anisotropic interlayer exchange observed in $CaMnTeO_6$, compared to the Sr and Pb materials. $SrMnTeO_6$ shows lower magnetic frustration than the Ca system owing to the less effective interactions between the layers. Although the Mn-Mn distances in $PbMnTeO_6$ are similar to those in $SrMnTeO_6$, the Pb system with lone-pair electrons features the least competing intralayer and interlayer coupling attributable to the sizable difference between the ICOBI within and between layer pathways. This work lays the foundation for incorporating the quantum-chemical approach into novel states of matter research. This integration can be a powerful tool to gain deeper understanding of physical phenomena while advancing materials design and development with distinct functions for foreseeable information technologies.

## 4. Methods

Crystals of $CaMnTeO_6$ are prepared by flux growth. $CaCO_3$, $MnCO_3$, and $Te(OH)_6$ powder (molar ratio 1:1:3) were ground and pressed into a pellet. The pellet is then sent to a box furnace, heated at 625°C for 40 hours, followed by slow cooling. Orange hexagon plate-shaped crystals can be separated from the flux. Single crystal diffraction experiments were performed on $CaMnTeO_6$ using a Bruker D8 Venture diffractometer with Mo Kα radiation ($\lambda$ = 0.71073 Å), and a Photon 100 detector at $T$ = 100 K. Data processing (SAINT) and scaling (SADABS) were performed using the Apex3 software. A synchrotron XRD pattern of $CaMnTeO_6$ was collected using the 11-BM beamline at Advanced Photon Source, Argonne National Laboratory. Data were collected from well ground crystal of $CaMnTeO_6$ at $T$ = 295 K and $\lambda$ = 0.45789 Å. Full-potential linearized augmented plane wave (FP-LAPW) spin-polarized electronic structure calculations were performed with the WIEN2k code.[31] The exchange and correlation energies were treated with Perdew−Burke−Ernzerhof generalized gradient approximation.[32] Pseudo-potential DFT calculations were calculated with Quantum Espresso (QE)[25] with the Generalized Gradient Approximation (GGA+U) of the exchange-correlation potential with the PBEsol parametrization[33], and the resulting wavefunctions and eigenvalues are used as the input for LOBSTER[26] for the DOS and bonding analysis[28a]. Heisenberg exchange parameters (*J*) were calculated using the Green's functions formalism[30a] with QE and the exchanges[30b] software. DC magnetization measurements on $CaMnTeO_6$ powder were performed with the vibrating sample magnetometer (VSM) option of Quantum Design Physical Properties Measurement System (PPMS) between 2 K ≤ $T$ ≤ 400

K at 0 T ≤ $\mu_0 H$ ≤ 7 T. Neutron diffraction was measured on 5 g sample of well grind powder on the HB-2A powder diffractometer at ORNL's HFIR reactor[34], measurements were taken with wavelengths of 2.41 A and 1.54 A in the temperature range 1.5 – 25 K. The low-temperature heat capacity was measured on a single crystal using a Quantum Design PPMS equipped with a dilution refrigerator.

**Supporting Information**

Supporting Information is available from the Wiley Online Library or from the author.

**Acknowledgement**

The work at Clemson University was supported by Arnold and Mabel Backman Foundation as a 2023 BYI grant to T.T.T. The neutron experiment at the Oak Ridge National Laboratory was in part funded by the National Science Foundation under award NSF-OIA-2227933. Participation of E.A. was supported by the NSF-CHE-2050042. We greatly appreciate the Halasyamani group for the SHG measurements. The research at Gdańsk University of Technology was supported by the National Science Centre (Poland) under SONATA-15 grant (UMO- 2019/35/D/ST5/03769). The work at the University of Utah was supported by an NSF Career Award (DMR-2145832). A portion of this research used resources at the High Flux Isotope Reactor, a DOE Office of Science User Facility operated by the Oak Ridge National Laboratory. Use of the Advanced Photon Source at the Argonne National Laboratory was supported by the U. S. Department of Energy, Office of Science, Office of Basic Energy Sciences, under Contract No. DE-AC02-06CH11357. This manuscript has been authored by UT-Batelle, LLC, under contract DE-AC05-00OR22725 with the US Department of Energy (DOE). The US government retains and the publisher, by accepting the article for publication, acknowledges that the US government retains a nonexclusive, paid-up, irrevocable, worldwide license to publish or reproduce the published form of this manuscript, or allow others to do so, for US government purposes. DOE will provide public access to these results of federally sponsored research in accordance with the DOE Public Access Plan (http://energy.gov/downloads/doe-public-access-plan). The work of L.N. and M.M. at G.T. (single-crystal thermomagnetic measurements) was funded by the U.S. Department of Energy, Office of Basic Energy Sciences, Materials Sciences and Engineering Division under Award DE-SC-0018660.